# An Anisotropic Constitutive Relationship by a Series of 8 Chain Models

Libin Yang

School of Industrial Automation, Beijing Institute of Technology, Zhuhai, 519088, China. E-mail: 12261@bitzh.edu.cn

Teng Long

Department of Mechanical and Materials Engineering, University of Cincinnati, Cincinnati, OH 45221. E-mail: longt8@mail.uc.edu

Lixiang Yang

Micro Phymicine LLC. West Chester OH 45069. E-mail: lxyoung12@gmail.com

**Abstract:** Hyperelastic models have been widely used to model polymers and soft tissues. However, most hyperelastic models are phenomenological material models. Based on statistical mechanics and molecular chain configuration, 8 chain model or Arruda-Boyce model is a physical model which can be used to understand how microstructures of chains affect macroscopic mechanical properties of polymers and soft tissues. Mechanical properties of many polymers and soft tissues are directional dependent. Polymer matrix can be reinforced by fibers. For soft tissues, ligaments and tendons will lead to anisotropic properties. Since matrix and reinforcements are composed of similar microstructural molecular chains, they can be modeled by using the same mathematical model. In this paper, a series of 8 chain models is used to understand composite properties. That is, an isotropic 8 chain model will be used to model matrix and anisotropic 8 chain models will be used to model fibers. Replacing $I_1$ in isotropic 8 chain model with $I_4$ in anisotropic 8 chain model is physically corresponding to changing representative 8 chain cubic cell to 8 chain slender cell. This treatment not only simplifies exist anisotropic mathematical structures but also keeps microscopic physics of 8 chain model unchanged.

## 1 Introduction

Hyperelastic constitutive models are important for industrial applications involving rubbers, elastomers, and other soft elastic materials [1] [2] [3] [4] [5] [6] [7]. Hyperelastic material models can be classified as compressible or incompressible depending on the importance of volume changes [8] [9] [10] [11] [12]. They can also be classified as phenomenological or physical, with the latter based on the underlying microstructure of the material [13] [14] [15] [16] [17]. They can be combined with equation of motion to model stress wave propagation in different solid materials [18] [19] [20] [21] [22]. The 8-chain or Arruda-Boyce model is one of the mostly widely used physical constitutive models in hyperelasticity [23] [24] [25] [26]. It can also be extended to include orthotropic mechanical behavior, with application in biotissue material characterization [27] [28] [29] [30] [31] [32]. It considers the microstructure of rubbery materials by statistically approximating the random distribution of polymer chains as 8 chains in a small representative cubic space. This 8-chain model has been successfully applied to a wide range of rubbery materials [33] [34] [35], and is widely used as a nonlinear spring component in other polymer constitutive models [36] [37] [38] [39] [40] [41] to capture nonlinear viscoelastic and rate-dependent viscoplastic behavior. However, due to the mathematical complexity of the inverse Langevin function in the Arruda-Boyce model, approximate forms are generally sought when numerically implementing it in finite-element codes [42] [43]. Currently, a short 5-term approximation is used in Abaqus and Ls-Dyna. While acceptable for small and moderate deformations, the 5-term approximation may be insufficient for large deformations.

On the other hand, 8-chain model is better than 3-chain model and 4-chain model. In 8-chain model, a representative cell is constructed which is shown in Fig. 1. Each side of the cell is assumed to be equal. There are 8 chains inside this cell. Each molecular chain starts from the center of cell and connect to each corner of the cell. Force and displacement relationship of each molecular chain is mathematically related by inverse Langevin function. For many polymer materials, molecular chains are cross linked and randomly distributed. If entanglements are neglected, these randomly distributed cross linked molecular chains are perfectly represented by 8 chain cells. It is well known that the same parameters in 8 chain model can be used to fit all simple tensile, bi-axial tensile, and shear test data. In 8 chain model, $\lambda_{chain} = \frac{1}{\sqrt{3}}(\lambda_1^2 + \lambda_2^2 + \lambda_3^2)^{1/2} = \frac{1}{\sqrt{3}}(I_1)^{1/2}$, where $I_1$ is the first stretch invariant. Elongation of each modular chain in the cell will be related to the change of $I_1$. The physical meaning of $I_1$ is average elongation of all line segments or all molecular chains in the polymer materials. $I_1$ is independent of coordinate system. In continuum mechanics, $I_1$ can be viewed as a function of the right Cauchy-Green deformation tensor $C$. When $C$ changes, $I_1$ will change as well. Since $C$ is independent of rigid body rotation, it is widely used in hyperelastic model to build constitutive relationship for large elastic deformations. Matrix $C$ only depends on the deformation of materials. It will be changed from one coordinate system to another. But $I_1$ will remain the same in different coordinate systems. Therefore, 8 chain model assumption connects polymer physics, average molecular chain elongation, with continuum mechanics hyperelastic mathematical structures.

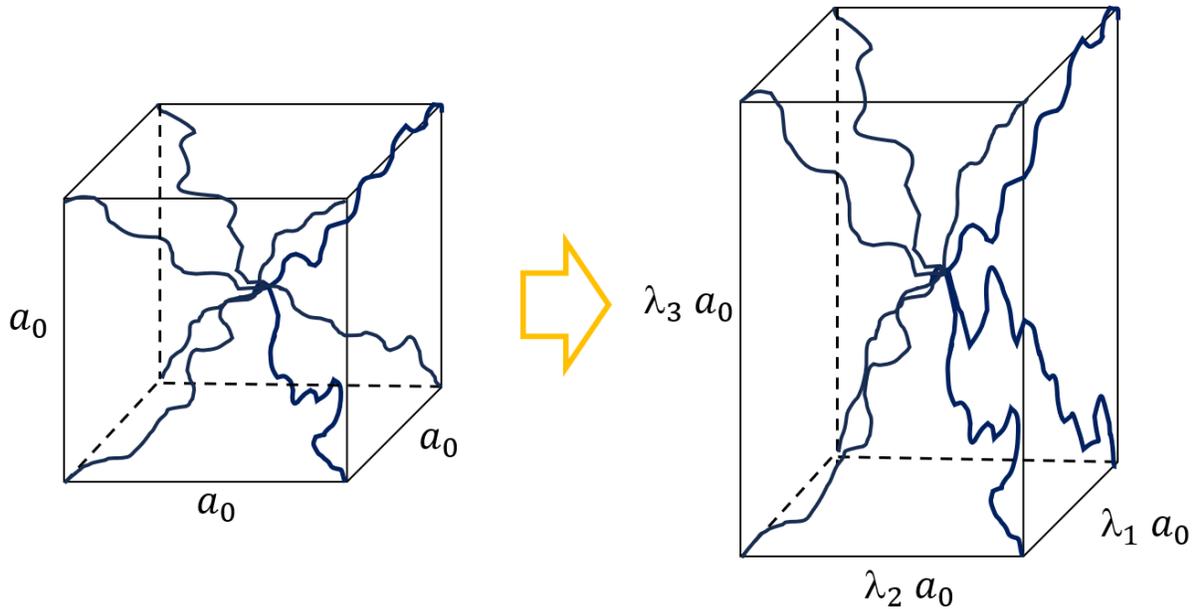

Fig. 1. A representative 8-chain cell. Undeformed cell (left) and deformed cell (right). The cubic cell in undeformed state is changed to a rectangular prism.

In continuum mechanics, a common way to add anisotropic properties into the hyperelastic model is to break the symmetry of strain energy function. Most researchers added $I_4$ into the strain energy function, where $I_4 = \boldsymbol{a_0} \cdot \boldsymbol{C a_0}$ and $\boldsymbol{a_0}$ is a unit vector that gives the orientation of the fibers through the materials. Aided by $I_4$, strain energy function will be decomposed into two parts. One part is isotropic strain energy function, and the other part is anisotropic strain energy function. For example, Holzapfel et al. [44] built an anisotropic hyperelastic model to nonlinear elastic behavior of arterial walls. In their constitutive model, neo-Hookean model is used for isotropic strain energy function and an exponential function of $I_4$ is used to describe the energy stored in the collagen fibers. Their model was extended by other researchers [45] and widely used to understand mechanical behavior of soft tissues [46, 47]. They claimed the model is based on the histological structure of arterial walls. But it should be considered as a phenomenological model. Especially, for neo-Hookean model, it can be derived based on Gaussian chain distribution. But from statistical mechanics perspective, it is based on microcanonical ensemble assumption [48]. That is why neo-Hookean model usually can only fit very small deformation experimental data. It will deviate from tests if deformation of materials is large.

Bischoff et al. [31] extended isotropic 8-chain model to an orthotropic 8-chain model. In their constitutive model, strain energy is decomposed into entropy part, repulsive part and bulk part. The freely jointed chain based inverse Langevin function is adopted for the mechanical behavior of a single chain. An orthotropic cell which has 8 chains is proposed to explain different mechanical properties in three principal directions. The total strain energy stored in the materials is the summation of the energy storage by an individual molecule chain which is calculated based on $I_4$ assumption. Their derivation is started from 8-chain model. After introducing many new concepts and new defined equations, their model is quite different from the original 8 chain or Arruda-Boyce model. Their mathematical structure, therefore, becomes unnecessarily hard.

In this paper, a new anisotropic model is proposed based on a series of 8-chain models. By keeping 8-chain mathematical structures, a series of 8-chain models will be adopted for both matrix and fibers. We think 8-chain model by itself is good enough to describe the anisotropic properties of soft materials. In another word, the 8 chain model will be used for matrix as well as reinforced fibers. If there is only one directional fiber, 8 chain model will be used for 2 times. If there are two directional fibers, 8 chain model will be used for 3 times. One for matrix, and two for fibers. The detail discussion will be given in section 5.

The paper is organized as follows. In the section 2, fundamental of continuum mechanics will be discussed. In the section 3, the accuracy of the Arruda-Boyce model at large elastic deformations by extending the current 5-term approximation to more terms will be illustrated. In the section 4, high order term Arruda-Boyce model was compared to uni-axial tension, bi-axial tension, and simple shear test results. In the following section, a series of 8-chain models will be used to model directional dependent materials. Conclusion will be given at the end of the paper.

## 2 Fundamentals of Continuum Mechanics

Any material model should agree with the thermodynamics second law. The entropy cannot decrease. In other word, it needs to agree with Clausius-Duhem inequality as shown below

$$- \rho^0(\dot{\psi} + \eta\dot{\theta}) + S : \dot{E} - \frac{Q \cdot \nabla\theta}{\theta} \geq 0 \tag{1}$$

where ψ, ρ, η, θ, S, E, and Q are Helmholtz free energy, density, entropy density, temperature, 2nd Piola-Kirchhoff stress tensor, Green Lagrange strain tensor, and heat flux in reference coordinate [49]. $\nabla\theta$ is the temperature gradient in initial configuration.

The constitutive equations of thermo-visco-elastic sold are introduced as

$$\psi(X,t) = \psi(E, \dot{E}, \theta, \dot{\theta}, \nabla\theta, X) \tag{2}$$

$$\eta(X,t) = \eta(E, \dot{E}, \theta, \dot{\theta}, \nabla\theta, X)$$

$$S(X,t) = S(E, \dot{E}, \theta, \dot{\theta}, \nabla\theta, X)$$

$$Q(X,t) = Q(E, \dot{E}, \theta, \dot{\theta}, \nabla\theta, X)$$

Substituting equation (2) in (1), the time rate of Helmholtz free energy $\psi$ is shown in Eqn. (3) and the Clausius-Duhem inequality can be finally arrange to Eqn. (4) and then Eqn. (5).

$$\dot{\psi} = \frac{\partial \psi}{\partial E} : \dot{E} + \frac{\partial \psi}{\partial \dot{E}} : \ddot{E} + \frac{\partial \psi}{\partial \theta} \dot{\theta} + \frac{\partial \psi}{\partial \dot{\theta}} : \ddot{\theta} + \frac{\partial \psi}{\partial \nabla\theta} : \nabla\dot{\theta} \tag{3}$$

$$-\rho^0 \left( \frac{\partial \psi}{\partial E} : \dot{E} + \frac{\partial \psi}{\partial \dot{E}} : \ddot{E} + \frac{\partial \psi}{\partial \theta} \dot{\theta} + \frac{\partial \psi}{\partial \dot{\theta}} : \ddot{\theta} + \frac{\partial \psi}{\partial \nabla\theta} : \nabla\dot{\theta} + \eta\dot{\theta} \right) + S : \dot{E} - \frac{Q \cdot \nabla\theta}{\theta} \geq 0 \tag{4}$$

$$-\rho^0 \left( \frac{\partial \psi}{\partial \dot{E}} : \ddot{E} + \frac{\partial \psi}{\partial \theta} \dot{\theta} + \frac{\partial \psi}{\partial \dot{\theta}} : \ddot{\theta} + \frac{\partial \psi}{\partial \nabla\theta} : \nabla\dot{\theta} + \eta\dot{\theta} \right) + \left( S - \rho^0 \frac{\partial \psi}{\partial E} \right) : \dot{E} - \frac{Q \cdot \nabla\theta}{\theta} \geq 0 \tag{5}$$

To maintain the Clausius-Duhem inequality valid, it is necessary to force $\frac{\partial \psi}{\partial \dot{E}}$, $\frac{\partial \psi}{\partial \dot{\theta}}$, and $\frac{\partial \psi}{\partial \nabla\theta}$ to be zero shown in Eqn. (6) to be zero, since we do not have $\ddot{E}$, $\ddot{\theta}$, and $\nabla\dot{\theta}$ in Eqn. (2)

$$\frac{\partial \psi}{\partial \dot{E}} = \frac{\partial \psi}{\partial \dot{\theta}} = \frac{\partial \psi}{\partial \nabla\theta} = 0 . \tag{6}$$

Finally, the Clausius-Duhem inequality is expressed as

$$-\rho^0 \left( \frac{\partial \psi}{\partial \theta} + \eta \right) \dot{\theta} + \left( S - \rho^0 \frac{\partial \psi}{\partial E} \right) : \dot{E} - \frac{Q \cdot \nabla\theta}{\theta} \geq 0. \tag{7}$$

In other word, the expression inside brackets on the left-hand side of Eq. (7) is the dissipated part and they can be expressed as

$$\eta^d = \frac{\partial \psi}{\partial \theta} + \eta \tag{8}$$

$$S^d = S - \rho^0 \frac{\partial \psi}{\partial E} \tag{9}$$

with

$$\eta = -\frac{\partial \psi}{\partial \theta} + \eta^d \tag{10}$$

$$S = \rho^0 \frac{\partial \psi}{\partial E} + S^d \tag{11}$$

The strain energy density function is written as

$$W = \rho^0 \psi. \tag{12}$$

In the continuous medium, vector *X* describes the material points initial configuration, while vector *x* is the current configuration. Their relation is shown as x= x(X, t) where at t=0 means initial configuration equal current configuration. The deformation gradient tensor is introduced as

$$F = \frac{\partial x}{\partial X} \tag{13}$$

where F is a two-point tensor and it relates the initial configuration and current configuration.

The right Cauchy-Green deformation tensor (**C**) and Green-Lagrange strain tensor (**E**) are defined as

$$C = F^T F, \quad E = \frac{1}{2}(C-I) \tag{14}$$

where they are positive definite and symmetric tensors.

To construct the constitutive model, we need coordinate measurement to be frame-indifference. Thus, the invariants are needed, and they are constructed as

$$I_1 = \text{tr}(C),\ I_2 = \frac{1}{2}(\text{tr}(C)^2 - \text{tr}(C^2)),\ I_3 = \det(C),\ I_4 = a_0 \cdot Ca_0,\ I_5 = a_0 \cdot C^2 a_0 \qquad (15)$$

where tr(C) and det(C) is the trace and the determinate of right Cauchy-Green deformation tensor. $a_0$ is the unit vector in fiber direction. It is noted that the det(C) also equals the volume fraction $V/V_0$. Their derivative with respect to right Cauchy-Green deformation tensor are [50]

$$\frac{\partial I_1}{\partial C} = I,\ \frac{\partial I_2}{\partial C} = I_1 I - C,\ \frac{\partial I_3}{\partial C} = I_3 C^{-1},\ \frac{\partial I_4}{\partial C} = a_0 \otimes a_0,\ \frac{\partial I_5}{\partial C} = a_0 \otimes Ca_0 + a_0 C \otimes a_0. \qquad (16)$$

Accordingly, the isochoric part of deformation gradient tensor and right Cauchy-Green deformation tensors are $\bar{F} = J^{-\frac{1}{3}} F,\ \bar{C} = J^{-\frac{2}{3}} C$.

Based on this definitions, modified invariants, e.g., $\bar{I}_1, \bar{I}_2, \bar{I}_3$, can be defined as

$$\bar{I}_1 = tr(\bar{C}),\ \bar{I}_2 = \frac{1}{2}(tr(\bar{C})^2 - \text{tr}(\bar{C}^2)),\ \bar{I}_3 = det(\bar{C}), \qquad (17)$$

A general anisotropic strain energy function can be written as

$$\psi = \psi_{iso}(\bar{I}_1, \bar{I}_2) + \psi_{aniso}(I_4, I_5) + \psi_{vol}(J), \qquad (18)$$

where $\psi_{iso}(\bar{I}_1, \bar{I}_2)$ is isotropic part, $\psi_{aniso}(I_4, I_5)$ is anisotropic part, and $\psi_{vol}(J)$ is volumetric part of strain energy function.

## 3 Approximation of 8-Chain Model

In 8 chain hyperelastic model, the single chain is described by the inverse Langevin function. Because of its mathematic complexity, the inverse Langevin function is approximated in several ways, such as the Pade approximation and the Warner approximation [51]. However, a key drawback of these approximations is the presence of singularities, which can cause numerical issues when these approximations are implemented in finite-element codes. To avoid singularities, we employ the Taylor expansion of inverse Langevin function, which is shown as

$$y = \mathcal{L}^{-1}(x) = \sum_{n=0}^{\infty} B_n x^n. \qquad (19)$$

In Eq. (19), all even coefficients of the Taylor expansion are zero. The odd coefficients of the Taylor expansion can be computed through recurrence relations [52]. The recurrence relations are used here to calculate the first 36 odd coefficients, which are provided in Table 1.

Table 1. Taylor series coefficients of the inverse Langevin function

| n | $B_n$ | n | $B_n$ |
|---|---|---|---|
| 1 | 3 | 3 | 1.8 |
| 5 | 1.6971 | 7 | 1.7589 |
| 9 | 1.8719 | 11 | 1.9972 |
| 13 | 2.1128 | 15 | 2.2023 |
| 17 | 2.2530 | 19 | 2.2558 |
| 21 | 2.2064 | 23 | 2.1064 |
| 25 | 1.9644 | 27 | 1.7962 |
| 29 | 1.6251 | 31 | 1.4799 |
| 33 | 1.3929 | 35 | 1.3953 |
| 37 | 1.5125 | 39 | 1.7581 |
| 41 | 2.1281 | 43 | 2.5962 |
| 45 | 3.1104 | 47 | 3.5933 |
| 49 | 3.9473 | 51 | 4.0638 |
| 53 | 3.8387 | 55 | 3.1924 |
| 57 | 2.0955 | 59 | 0.6016 |
| 61 | -1.0928 | 63 | -2.4914 |
| 65 | -2.2916 | 67 | 3.3051 |
| 69 | 26.2421 | 71 | 105.5569 |

The strain energy function associated with the Arruda-Boyce model [53] can be expressed as

$$U = nk\Theta N \left( \frac{\lambda_{\text{chain}}}{\sqrt{N}} \beta + \ln \frac{\beta}{\sinh \beta} \right) - \Theta c', \qquad (20)$$

where $n$ is the chain density, $k$ is Boltzmann's constant, $\Theta$ is temperature, $N$ is the number of rigid links of equal chain length $l$. $c'$ is a combination of constants, and

$$\beta = \mathcal{L}^{-1} \left[ \frac{\lambda_{\text{chain}}}{\sqrt{N}} \right], \qquad \lambda_{\text{chain}} = \frac{1}{\sqrt{3}} \left( \lambda_1^2 + \lambda_2^2 + \lambda_3^2 \right)^{1/2},$$

$\mathcal{L}^{-1}(*)$ represents the inverse Langevin function. The chain which force and displacement relationship can be expressed as inverse Langevin function is called inverse Langevin chain. Due to the complex mathematical form of the inverse Langevin function, expansionist Taylor series approximation is used in finite-element software such as Abaqus and Ls-Dyna for simplification. However, the approximated expansion of the Arruda-Boyce model never goes beyond the first 5 terms in FEA code. For instance, the hyperelastic Arruda-Boyce potential from the Abaqus 2021 theory guide has the following form:

$$W = nk\Theta \sum_{i=1}^{5} \frac{C_i}{N^{i-1}} (\bar{I}_1^i - 3^i) + \frac{1}{D}\left(\frac{J^2 - 1}{2} - \ln J\right), \tag{21}$$

where

$$C_1 = \frac{1}{2}, \quad C_2 = \frac{1}{20}, \quad C_3 = \frac{11}{1050}, \quad C_4 = \frac{19}{7000}, \quad C_5 = \frac{519}{673750}$$

and $J = det(F)$. $\bar{I}_1$ is the deviatoric first strain invariant.

The 5-term approximation of the Arruda-Boyce is reasonably accurate for small and moderate deformations, but will deviate from true solution at large deformations, practically for small locking stretch or when actual stretch is larger than locking stretch. It will be shown in the following examples. To improve computation accuracy, the 5 terms approximation is extended to more terms which can be used by Abaqus. A brief derivation will lead to this. First, by taking the derivative of Eq. (20) with respect to stretch and enforcing incompressibility, which can be expressed as

$$\frac{dU}{d\lambda_i} = \frac{nk\Theta}{3}\sqrt{N}\mathcal{L}^{-1}\left[\frac{\lambda_{chain}}{\sqrt{N}}\right]\frac{\lambda_i}{\lambda_{chain}} \tag{22}$$

where

$$\lambda_{chain} = \frac{1}{\sqrt{3}}(\lambda_1^2 + \lambda_2^2 + \lambda_3^2)^{1/2}.$$

The principal Cauchy stress difference can be obtained as

$$T_1 - T_2 = \lambda_1 \frac{dU}{d\lambda_1} - \lambda_2 \frac{dU}{d\lambda_2} = \frac{nk\Theta}{3} \sqrt{N} \mathcal{L}^{-1}\left[\frac{\lambda_{chain}}{\sqrt{N}}\right] \frac{\lambda_1^2 - \lambda_2^2}{\lambda_{chain}}. \qquad (23)$$

Next, the approximation of inverse Langevin function from Section 2 is substituted into Eq. (22), and integration can be performed term by term using the Taylor expansion of the inverse Langevin function. The first 36 nonzero coefficients of Taylor expansion of the Arruda-Boyce model are shown in Table 2. Note that the first five coefficients in Table 2 are identical to those used in Abaqus (cf. Eq. (21)).

Table 1. Taylor series coefficients of the Arruda-Boyce model

| n | $C_n$ | n | $C_n$ |
|---|---|---|---|
| 1 | 0.5 | 3 | 0.05 |
| 5 | 0.0105 | 7 | 0.0027 |
| 9 | 7.7032e-4 | 11 | 2.281e-4 |
| 13 | 6.9006e-5 | 15 | 2.0979e-5 |
| 17 | 6.3590e-6 | 19 | 1.9101e-6 |
| 21 | 5.6615e-7 | 23 | 1.6515e-7 |
| 25 | 4.7388e-8 | 27 | 1.3412e-8 |
| 29 | 3.7751e-9 | 31 | 1.0744e-9 |
| 33 | 3.1724e-10 | 35 | 1.0004e-10 |
| 37 | 3.4245e-11 | 39 | 1.2605e-11 |
| 41 | 4.8440e-12 | 43 | 1.8803e-12 |
| 45 | 7.1823e-13 | 47 | 2.6505e-13 |
| 49 | 9.317e-14 | 51 | 3.0745e-14 |
| 53 | 9.3222e-15 | 55 | 2.4920e-15 |
| 57 | 5.2644e-16 | 59 | 4.8702e-17 |
| 61 | -2.8537e-17 | 63 | -2.1008e-17 |
| 65 | -6.2457e-18 | 67 | 2.9144e-18 |
| 69 | 7.4930e-18 | 71 | 9.767e-18 |

Arruda-Boyce model is based on a single ideally rotational chain and eight-chain configuration assumptions. For each chain, bond between each monomer is assumed to be rigid and free to rotate. Chain length is constant. Initial entire chain length is given as $\sqrt{N}l$. Fully stretched entire chain length is $Nl$. Therefore, the locking stretch is fully stretched entire chain length, e.g., $Nl$, divided by initial entire chain length, e.g. $\sqrt{N}l$, which is $\sqrt{N}$. When stretch approaches to this locking stretch value, more Taylor series coefficients are needed to capture inverse Langevin function. In

Fig. 2, we plotted Arruda-Boyce model in uniaxial tension with maximal stretch less than locking stretch with different terms Taylor expansion approximation.

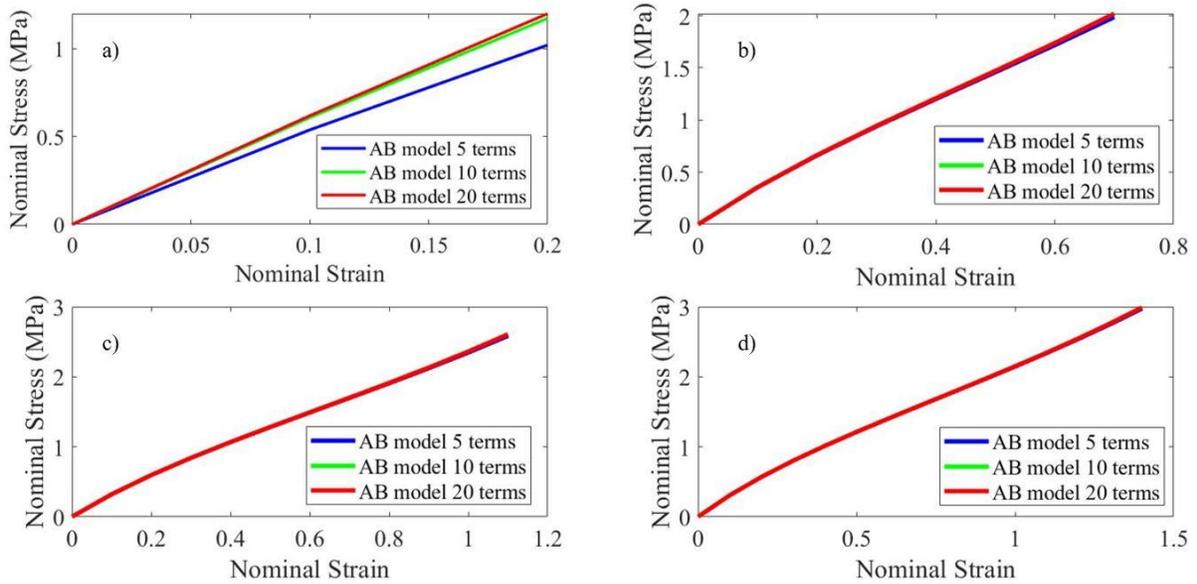

Fig. 2. Arruda-Boyce model is plotted for uniaxial tensile with $nk\Theta = 1$ for simplicity. Locking stretch $\sqrt{N}$ is set to different numbers a) 1.22, b) 1.73, c) 2.12, d) 2.45. (Nominal strain = stretch−1). Large difference is shown in case a where 10 and 20 terms Taylor expansion diverges from 5 terms Taylor expansion. But in the other cases, their difference is small and negligible.

Recently, in order to understand polymer fatigue, toughness and fracture [54] [55] [56], the constant bond length assumed in Arruda-Boyce model is relaxed. The bonding length between monomers can be stretched [57] [58] [59]. This means the stretch can even be larger than the locking stretch. Therefore, we need to understand the approximation of the stretch larger than locking stretch. In Fig. 3, Arruda-Boyce model is plotted when a single chain is stretched even larger than locking length. With more terms adding into inverse Langevin function, Arruda-Boyce model shows a big difference when stretch is larger than locking length.

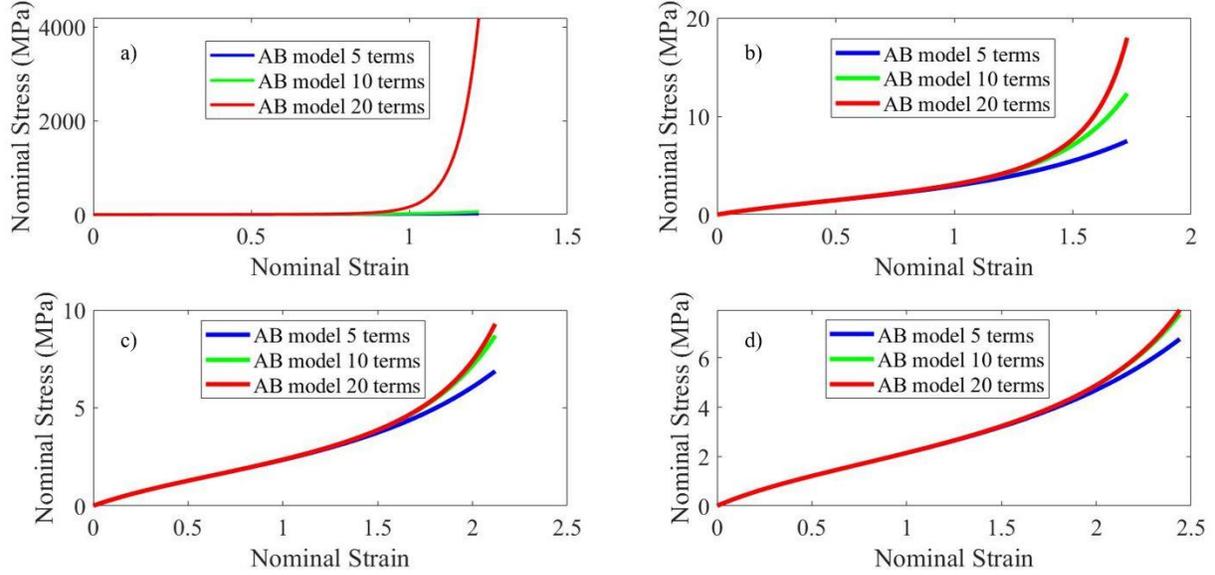

Fig. 3. Arruda-Boyce model is plotted for uniaxial tensile with nkΘ = 1 for simplicity. Locking stretch $\sqrt{N}$ is set to different numbers a) 1.22, b) 1.73, c) 2.12, d) 2.45. But actual maximum stretch is set to be a) 2.22, b) 2.73, c) 3.12, d) 3.45. (Nominal strain = stretch−1). 10 and 20 terms Taylor expansion models diverge from 5 terms Taylor expansion model when actual stretch is larger than locking stretch.

**4 Comparison of Isotropic Arruda-Boyce Model with Experimental Data**

For uniaxial tension of Arruda-Boyce model, nominal stress, $T_{o1}$, is related to stretch, $\lambda$, by

$$T_{o1} = \frac{nk\Theta}{3} \frac{\sqrt{N}}{\lambda_{chain}} \mathcal{L}^{-1}\left[\frac{\lambda_{chain}}{\sqrt{N}}\right](\lambda - 1/\lambda^2), \qquad (24)$$

where $\lambda_{chain} = (\lambda^2 + 2/\lambda)^{1/2}/\sqrt{3}$.

For equi-biaxial extension of Arruda-Boyce model, nominal stress can be written as

$$T_{o1} = \frac{nk\Theta}{3} \frac{\sqrt{N}}{\lambda_{chain}} \mathcal{L}^{-1}\left[\frac{\lambda_{chain}}{\sqrt{N}}\right](\lambda - 1/\lambda^5), \qquad (25)$$

where $\lambda_{chain} = (2\lambda^2 + 1/\lambda^4)^{1/2}/\sqrt{3}$.

For pure shear or plane strain tension of Arruda-Boyce model, nominal stress can be written as

$$T_{o1} = \frac{nk\Theta}{3} \frac{\sqrt{N}}{\lambda_{chain}} \mathcal{L}^{-1}\left[\frac{\lambda_{chain}}{\sqrt{N}}\right](\lambda - 1/\lambda^3), \qquad (26)$$

where $\lambda_{chain} = (\lambda^2 + 1 + 1/\lambda^2)^{1/2}/\sqrt{3}$.

Eq.(24) to Eq.(26) can be plotted against experimental data by Teloar's classical experimental data in uniaxial tension, biaxial extension, and pure shear test data. 5 terms, 10 terms and 20 terms

Taylor expansions of inverse Langevin function are used in Eq. (24), Eq. (25), and Eq. (26). Parameters $N = 26.5$ and $\frac{nk\Theta}{3} = 0.09\ MPa$ are used for all uniaxial, shear, and bi-axial models. The comparison is shown in Fig. (4). For uniaxial tension test, 5 terms Taylor expansion model is accurate when stretch is up to 6. When stretch is larger than 6, more Taylor expansion terms are needed. For Biaxial loading case, 5 terms Taylor expansion model is good for stretch is up to 4.5. In the large stretch, for example, $\lambda > 4.5$, 10 or 20 terms of Taylor expansion are needed to match the experimental data. Shear model results are like uniaxial tension model results.

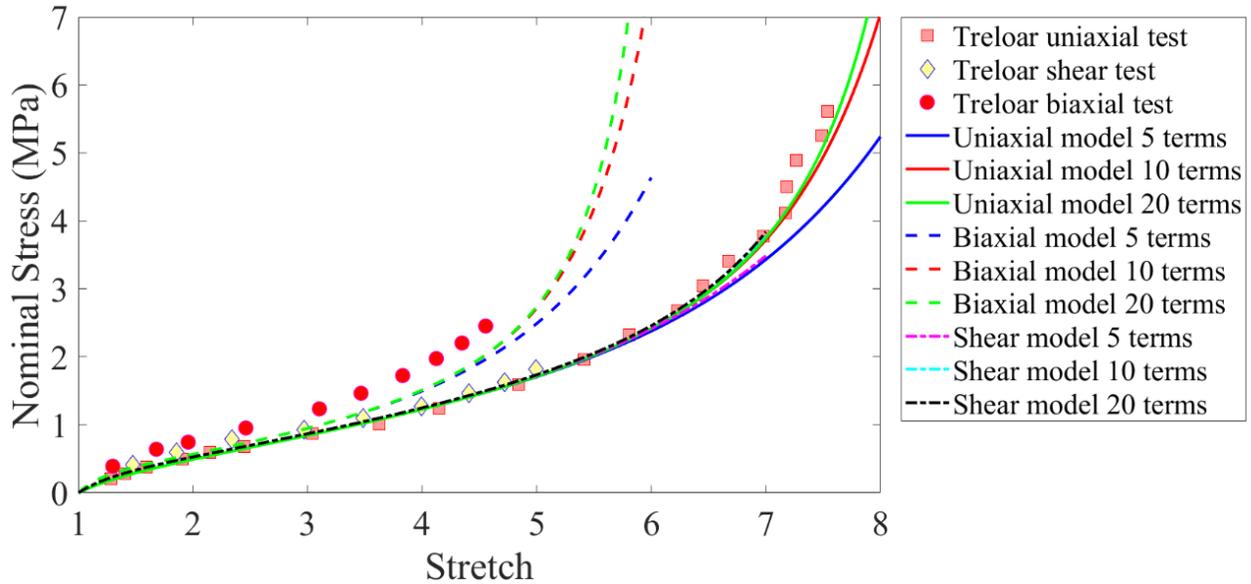

Fig. 4. Arruda-Boyce model is compared against classical Treloar uniaxial, shear, and biaxial test data. Arruda-Boyce model is plotted using 5 terms, 10 terms, and 20 terms Taylor expansion of inverse Langevin function. Parameters $N = 26.5$ and $\frac{nk\Theta}{3} = 0.09\ MPa$ are used for Arruda-Boyce model in uniaxial, shear, and biaxial cases.

To test if the 8-chain model can fit the real soft tissue experiment data, the uniaxial, biaxial, and pure shear experiments are used to fit the model. There are many available experiments from literature, however, we only select two cases to demonstrate the model for experiment data. The uniaxial and biaxial experiments are from human linea alba [60], while porcine livers contribute the pure shear parameter fit [61]. To find the material model parameter, the optimization method is used. The Genetic algorithm [62] is used to find the global solution. It is followed by Nelder–Mead method, a local optimization method, to refine the solution [29]. The Figure 5 below shows good fits between the model response and experiment data. The material model parameters are shown in Table 3. Those figures (a)-(c) are fitted by 20-term model. Other terms can also fit the experiment data. However, there are many combinations between different terms and experiments.

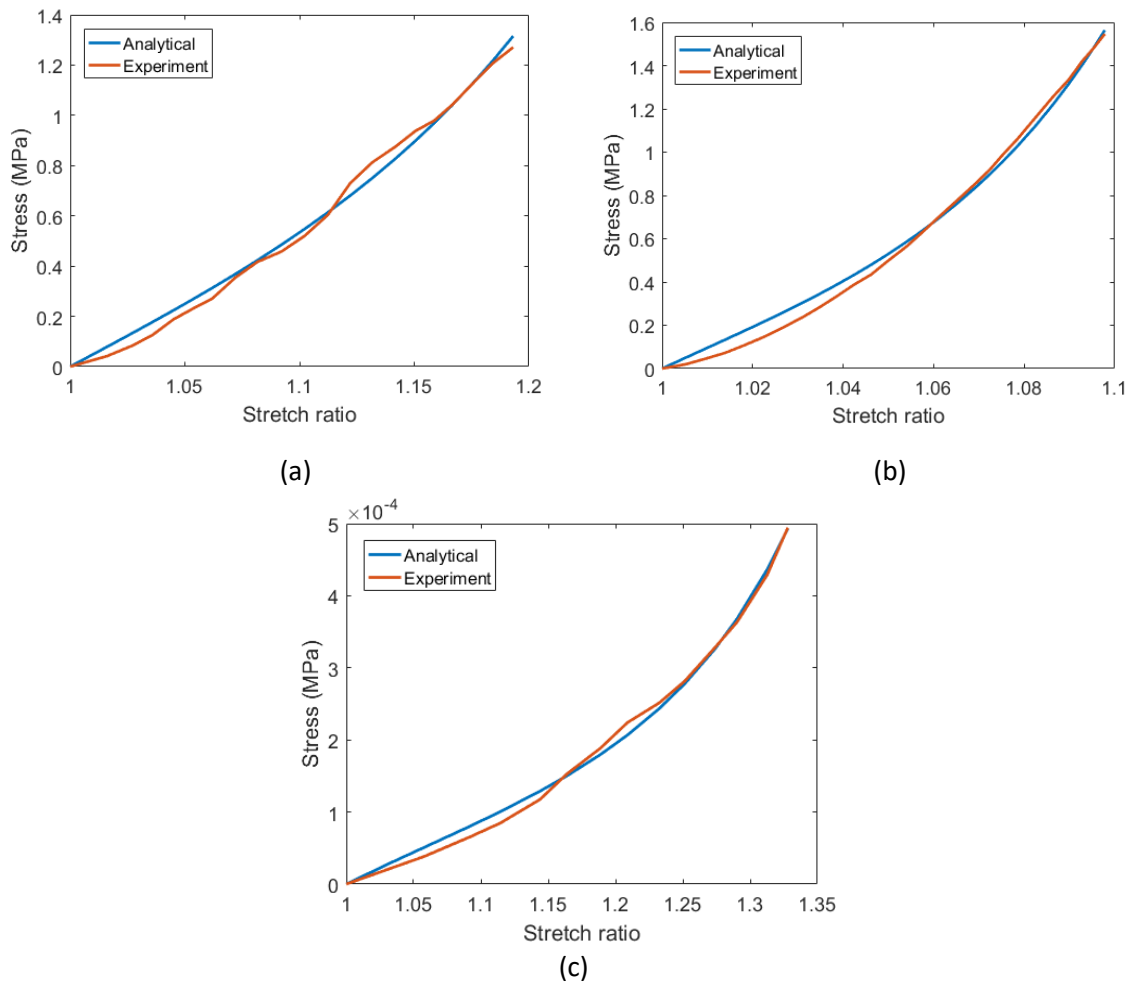

Fig. 5. The Cauchy stress-stretch response. (a) Uniaxial tension. (b) Biaxial tension. (c) Pure shear.

Table 3. The material model parameters of uniaxial, biaxial, pure shear test

| Item | $nk\Theta/3$ | N |
|---|---|---|
| Uniaxial | 1.837344E-1 | 1.067091 |
| Biaxial | 3.503356E-4 | 7440155E-1 |
| Pure shear | 4.394207E-5 | 1.1446898 |

## 5 An Anisotropic Model by a Series of 8-Chain Models

Currently, most mathematical structures of anisotropic hyperelastic models are based on the fourth invariant, e.g., $I_4$. If we take a look at anisotropic strain energy functions, additional strain energy stored by fibers is added to original isotropic part. After taking derivative with respect to strain tensor, an additional force term is added to stress tensor. This additional force term is due to fibers, mathematically, which can be modelled by the fourth invariant. Microscopically, these reinforced fibers, by themselves, are made of molecular chains. The randomness of these chains is not different from the matrix polymers. Therefore, both the matrix and fibers can be modeled by 8-chain model. The difference between matrix and reinforced fibers comes from their macroscopic geometry and microscopic chain density. By using different representative cell geometry, one Arruda-Boyce model can be added to another Arruda-Boyce model to find anisotropic properties of reinforced soft materials. In another word, A series of Arruda-Boyce models is used to find mechanical properties of reinforced soft materials such as polymers and soft tissues. In the past, Arruda-Boyce model was extended to be an orthotropic hyperelastic model [31]. But their idea was based on strain energy contribution by individual chains. Their mathematical structure is unnecessarily complex. A lot of new assumptions were introduced as well. The purpose of the series of 8-chain models is to keep the mathematical structures simple and microscopic physics of Arruda-Boyce model intact.

Idea of the series of 8-chain models is to use different representative cell for matrix materials and fibers. For matrix materials, there is no directional orientation. the representative cell element is cubic. Original Arruda-Boyce model is used. For fibers, the representative cell element is like fiber shape. As shown in Fig. 6, 8-chain structured cell will be used for both matrix and fibers. Because the molecular chains in the fiber cell have orientation preference, chain length in $a_0$ direction is much longer than the other two directions, e.g., $a_0 > b_0, c_0$. Under external force, the change of $b_0, c_0$ can be neglected compared to the change of $a_0$, as shown in Fig. 7. In isotropic 8-chain model, $\lambda_{chain} = \frac{1}{\sqrt{3}}(\lambda_1^2 + \lambda_2^2 + \lambda_3^2)^{1/2} = \frac{1}{\sqrt{3}}(I_1)^{1/2}$. Anisotropic 8-chain model can be obtained simply by replacing $\lambda_{chain}$ by $\lambda_{fiber\_chain} = \frac{1}{\sqrt{3}}(I_4 + 2)^{1/2}$ and keep its mathematical structure the same as the isotropic 8-chain model. Replacing $I_1$ by $I_4 + 2$ is physically related to change 8-chain representative cubic cell to 8-chain representative fiber slender cell, as shown in Fig. 6. The detail derivation will be demonstrated as below.

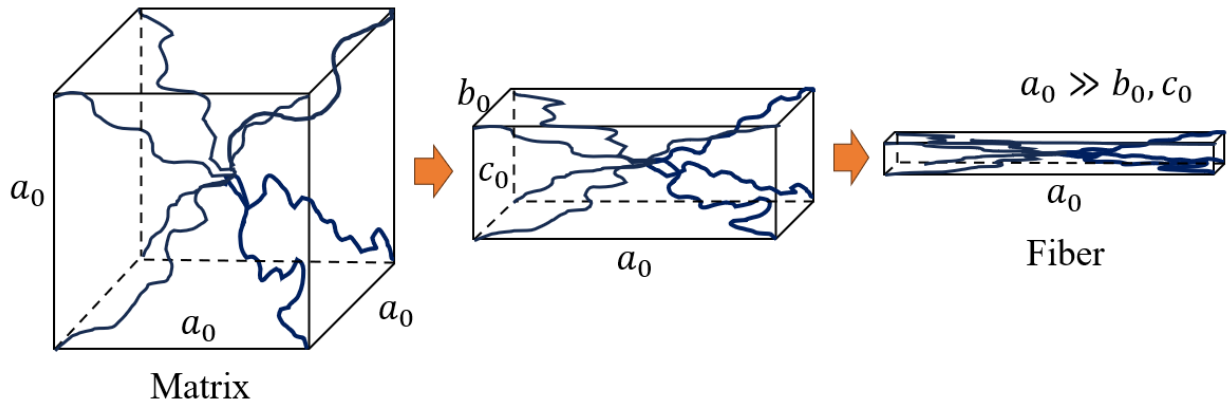

Fig. 6. Representative cell elements for matrix and fibers. Regular representative cell element is used for matrix materials. A slender cell element is used for fibers.

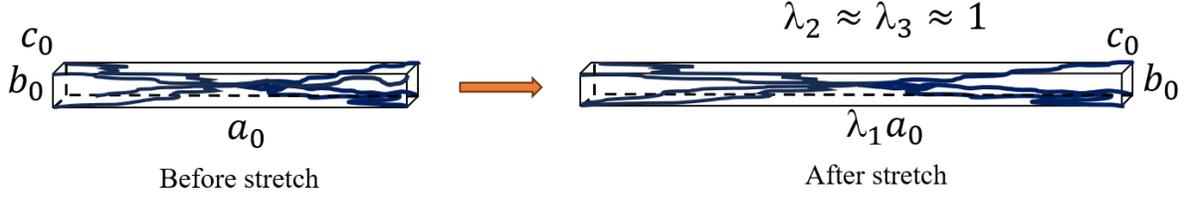

Fig. 7. A representative cell for fiber before stretch and after stretch. The stretches in $b_0$ and $c_0$ directions are assumed to be one.

Suppose there exist a fiber in an arbitrary direction respect to principal axis of matrix, as shown in Fig. 8. The unit vector $\boldsymbol{a_0}$ is defined in three-dimensional space by two angles $\varphi$ and $\theta$.

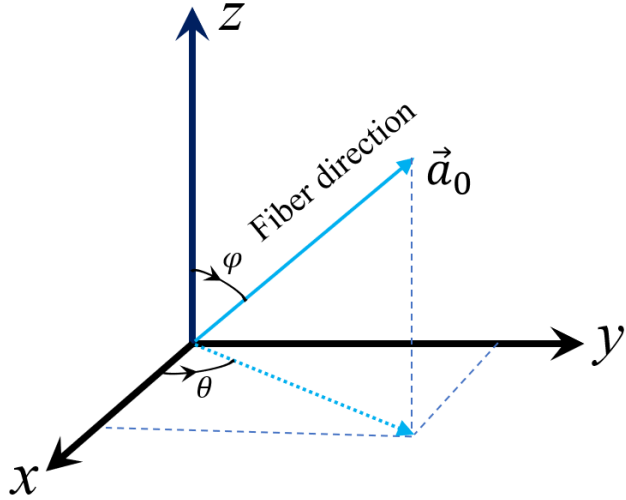

Fig. 8. Schematic sketch of fiber direction in 3-dimensional principal coordinate system

Mathematically the orientation of the fiber can be expressed as $\boldsymbol{a_0} = x\vec{\imath} + y\vec{\jmath} + z\vec{k}$ where $x = sin\varphi cos\theta, y = sin\varphi sin\theta, z = cos\varphi$. A tensor matrix of the unit vector can be defined as $a_0 \otimes a_0$ which is written explicitly as

$$a_0 \otimes a_0 = \begin{pmatrix} sin^2\varphi cos^2\theta & sin^2\varphi sin\theta cos\theta & sin\varphi cos\varphi cos\theta \\ sin^2\varphi sin\theta cos\theta & sin^2\varphi sin^2\theta & cos\varphi sin\varphi sin\theta \\ sin\varphi cos\varphi cos\theta & cos\varphi sin\varphi sin\theta & cos^2\varphi \end{pmatrix}. \quad (27)$$

If let $\varphi = 90^0$ and $\theta = 0^0$, tensor $a_0 \otimes a_0$ becomes

$$a_0 \otimes a_0 = \begin{pmatrix} 1 & 0 & 0 \\ 0 & 0 & 0 \\ 0 & 0 & 0 \end{pmatrix},$$

which means the fiber direction is parallel to the first principal direction, e.g., $x$ direction. The fourth invariant, e.g., $I_4$, is defined as $\mathbf{C} : (a_0 \otimes a_0)$. $\mathbf{C}$ is the right Cauchy-Green deformation tensor which is independent of rigid body rotation. In principal direction, the eigenvalues of matrix $\mathbf{C}$ is $\lambda_1^2, \lambda_2^2, \lambda_3^2$. It can be written in a matrix form

$$\mathbf{C} = \begin{pmatrix} \lambda_1^2 & 0 & 0 \\ 0 & \lambda_2^2 & 0 \\ 0 & 0 & \lambda_3^2 \end{pmatrix}.$$

Therefore, in principal direction, $I_4$ can be written as

$$I_4 = \lambda_1^2 \sin^2\varphi \cos^2\theta + \lambda_2^2 \sin^2\varphi \sin^2\theta + \lambda_3^2 \cos^2\varphi. \tag{28}$$

If $\varphi = 90^0$ and $\theta = 0^0$, $I_4 = \lambda_1^2$. If $\varphi = 90^0$ and $\theta = 90^0$, $I_4 = \lambda_2^2$. If $\varphi = 0^0$, $I_4 = \lambda_3^2$.

Recall 8-chain model, strain energy function is also expressed in the principal direction as a function of the first invariant, $I_1 = \lambda_1^2 + \lambda_2^2 + \lambda_3^2$. Since a fiber is a one-dimension structure, the length change of a fiber will not lead to much change of fiber cross section. For a slender fiber, symmetry of $I_1$ will be broken. Assuming a fiber is aligned in the $x$ direction, $\lambda_2^2 \approx \lambda_3^2 \approx 1$ and $I_1 \approx \lambda_1^2 + 2$. If a fiber is aligned in the $y$ direction, $I_1 \approx \lambda_2^2 + 2$. Similarly for $z$ direction, $I_1 \approx \lambda_3^2 + 2$. Inspired by this, adding 2 to Eq. (28), the definition of $I_4$, we obtain

$$I_4 + 2 = (\lambda_1^2 + 2)\sin^2\varphi \cos^2\theta + (\lambda_2^2 + 2)\sin^2\varphi \sin^2\theta + (\lambda_3^2 + 2)\cos^2\varphi, \tag{29}$$

where $\sin^2\varphi + \cos^2\varphi = 1$ and $\sin^2\theta + \cos^2\theta = 1$ are used. If a fiber is in $x$ direction, e.g., $\varphi = 90^0$ and $\theta = 0^0$. Eq. (29) becomes

$$I_4 + 2 = (\lambda_1^2 + 2) \approx I_1, \tag{30}$$

where the right-hand side is $I_1$ if representative 8 chain slender cell is assumed for fibers, as shown Fig.7. That is why isotropic 8 chain model will become anisotropic 8 chain model if $I_1$ is replaced by $I_4 + 2$. Physically, representative 8-chain cubic cell is changed to representative 8-chain slender cell because molecular chains are dominate in only one direction.

The total strain energy for matrix and fibers is given as

$$U_{total} = U_{matrix} + U_{fiber}$$
$$= nk\Theta N \left(\frac{\lambda_{chain}}{\sqrt{N}} \beta + \ln \frac{\beta}{\sinh \beta}\right) \quad (31)$$
$$+ n_{fiber} k\Theta N_{fiber} \left(\frac{\lambda_{fiber\_chain}}{\sqrt{N_{fiber}}} \beta_{fiber} + \ln \frac{\beta_{fiber}}{\sinh \beta_{fiber}}\right),$$

where $n_{fiber}$ is the fiber chain density, $N_{fiber}$ is the number of rigid links of equal fiber chain length and $\beta_{fiber} = \mathcal{L}^{-1}\left[\frac{\lambda_{fiber\_chain}}{\sqrt{N_{fiber}}}\right]$. Incompressible Cauchy stress is given as $T_i = \lambda_i \frac{dU_{total}}{d\lambda_i} - p^*$, where no sum on $i$ and $p^*$ is a constant depending on boundary conditions. Strain energy of fibers will depend on fiber orientation. If a fiber is in x-axis direction, it only depends on the principal stretch of $\lambda_1$. Therefore, additional stress term will be added into principal direction of x-axis. And stress on y-axis and z-axis direction will remain the same. If fibers are in arbitrary direction, strain energy of fibers will depend on orientation angles and principal stretches. Mechanical properties in the reinforced directions are usually dominated by the fibers. The anisotropic 8 chain model is plotted against two sets of experimental data. In Fig. 9a, the model is used to fit vitro uniaxial tensile tests of rabbit skin with lateral contraction constrained [63]. More tendons align with $x_1$ direction make stress in $x_1$ direction is stronger than stress in $x_2$ direction. Nominal stress will reach 17 kPa when stretch in $x_1$ direction is 1.5. Stretch in $x_2$ direction need be 1.9 to get the same stress level. The parameters used in anisotropic 8-chain model are shown in table 4a. Numerical nominal stress in $x_2$ direction is calculated by using parameters of matrix. Numerical nominal stress in $x_1$ direction is summation of stress contributed by matrix and fiber. Fig. 9b shows the comparison of model and test results of human linea alba by Cooney et al. [60]. The test results are the mean values of 14 samples. Strong stress is along with transverse direction. Stress response in longitudinal direction is weak. Therefore, numerical true stress in transverse direction is calculated by using matrix parameters and fiber parameters given in table 4b. Only matrix contribute the stress in longitudinal direction. All numerical calculations are performed by using Matlab. Numerical results can capture test data in different directions with reasonable accuracy.

Table 4a. The material parameters of anisotropic 8 chain model to fit rabbit skin

|  | $n$ (1/$m^3$) | N | $\Theta$ (K) | $k$ (J/K) |
|---|---|---|---|---|
| Matrix | 3.75× $10^{22}$ | 1.6 | 330 | $1.38 \times 10^{-23}$ |
| fiber | 1.2× $10^{22}$ | 1.39 | 330 | $1.38 \times 10^{-23}$ |

Table 4b. The material parameters of anisotropic 8 chain model to fit human linea alba

|  | $n$ (1/$m^3$) | N | $\Theta$ (K) | $k$ (J/K) |
|---|---|---|---|---|
| Matrix | 9.6× $10^{25}$ | 1.2 | 330 | $1.38 \times 10^{-23}$ |
| fiber | 39.2× $10^{25}$ | 1.152 | 330 | $1.38 \times 10^{-23}$ |

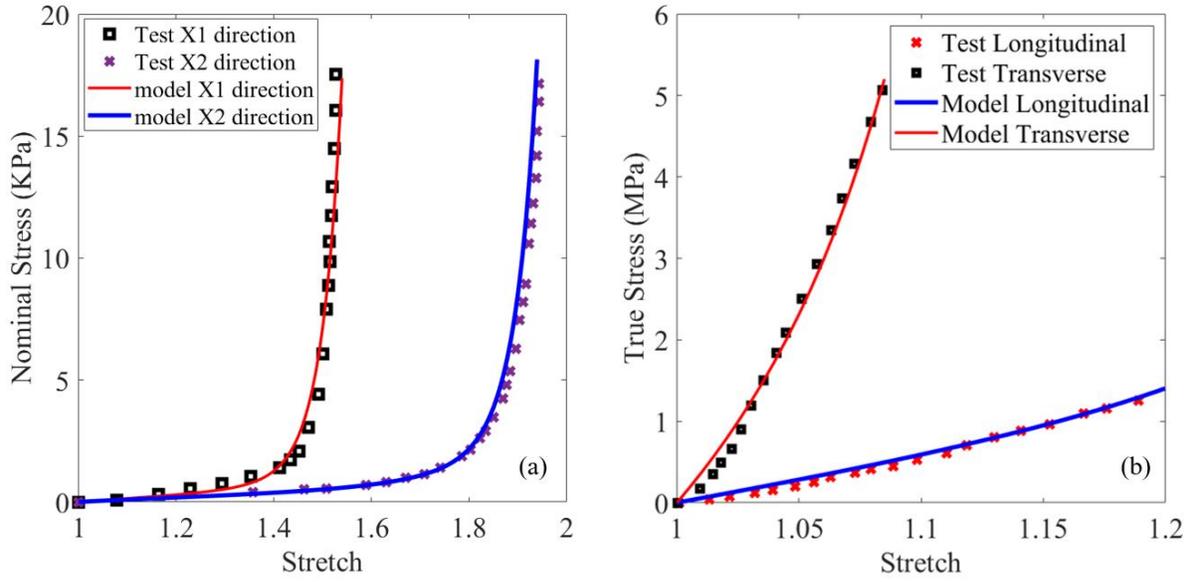

Fig. 9 A series of 8-chain models was used to fit test data (a) Test data is from uniaxial tests on rabbit skin by Lanir and Fung [63]. (b) Test data is from uniaxial tests on human linea alba by Cooney et al. [60]

## 6. Discussion

### 6.1 Geometric nonlinearity

Since no additional assumptions other than 8-chain model are used to derive the anisotropic 8-chain model, implementing it to finite element code such as Abaqus and Ansys has no difficulty. In finite element analysis, calculated results will become more understandable if there exist no geometric nonlinearity and material nonlinearity. However, both geometric nonlinearity and material nonlinearity are unavoidable. Entire continuum mechanics is treating these geometric nonlinearity and material nonlinearity. For small deformation problems, geometric nonlinearity is usually related to large rigid body rotation. When a new finite deformation constitutive model is built, rigid body rotation should be taken out from stress and strain. In the anisotropic 8-chain model, representative cell of fiber can be rotated to any direction. $\lambda_{\text{fiber\_chain}} = \frac{1}{\sqrt{3}}(I_4 + 2)^{1/2}$ is independent on rigid body rotation. For large deformation materials such as polymers and tissues, material nonlinearity need be captured by material models. That is why the force and stretch relationship of each molecular chain is highly nonlinear. By affine deformation assumption, mechanical behavior of macroscopic soft materials is nonlinear too. In finite element analysis, the computational domain is meshed to be a small size element. A constitutive material model is assigned to this small size element. If a local coordinate system is also assigned to the element, all initial conditions, boundary conditions and constitutive model need use this local coordinate system. For example, external force direction will change with the defined local coordinate system since the local coordinate system will rotate with each element. Principle directions of constitutive models will change with the local coordinate system of each element. Geometry nonlinearity due to large body rotation can be computed by relationship of local coordinate system and global coordinate system. Especially for Hooke's law which has small stress and strain definition, the rigid body rotation is not included in the material model.

## 6.2 Material Nonlinearity

For soft materials such as polymers and soft tissues, material nonlinearity is due to molecular chain response under an external force. For 8-chain model, force and displacement of a single molecular chain observe inverse Langevin function relationship by assumption of freely joint chain. There are some other mathematical force-displacement relationships such as worm like chain or semi-rigid chain model. Foundation of them is based on freely rotated chain assumption and entropic elasticity. However, a real molecular chain cannot be freely rotated. For example, Carbon to carbon connection, e.g., C-C bond, can be single bond, double bond and triple bond. As shown in Fig. 10, single bond is made up of $\sigma - bond$. Double bond is made up of $\sigma - bond$ and of $\pi - bond$. The constraint is higher if double bond is forced to rotate. To understand the energy of these bond rotations, Femi-Dirac function is necessary [48]. Femi-Dirac function describe the probability of occupied covalent electrons in each energy levels. Bond rotation of a molecular chain is related to energy level changes of occupied covalent electrons. Some development can be seen in our recent work [48, 39].

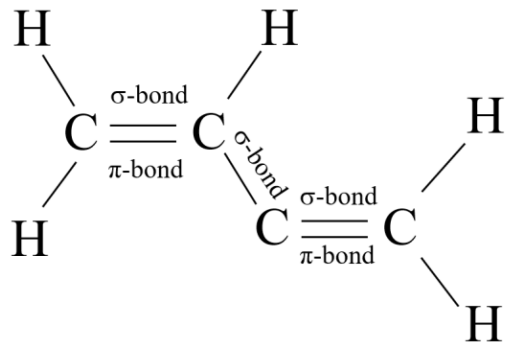

Fig.10 a schematic sketch of a polymer chain with different types of bonds. The double bonds are made up of $\sigma - bond$ and of $\pi - bond$. The single bond is only made up of $\sigma - bond$. Rotation of single bond and double bond is related to energy level changes of covalent electrons.

## 7 Conclusion

Soft materials such as polymers and soft tissues are made of random distributed molecular chains. The random cross linked molecular chains can be perfectly represented by 8 chain cells. Depending on chain orientations, cubic 8 chain cell is used for bulk matrix materials while slender 8 chain cell is used for fibers. Mathematical structure of 8 chain model is applied to both matrix and fiber. Strain energy of matrix is isotropic and strain energy of fibers will depend on fiber direction. Fiber will contribute additional force in the fiber direction. Different direction fibers can all be modeled as 8-chain model by fiber orientations controlled by two angles. Since no additional assumptions except 8 chain mathematical structures are used in the derivation, employment of the model for finite element analysis will be easy.

## 8 Funding

The author(s) received no financial support for the research, authorship, and/or publication of this article.